\documentclass[10pt]{article}
\usepackage{fullpage}
\usepackage{amsmath}
\usepackage{amssymb}
\usepackage[dvips]{epsfig}
\usepackage{color}

\def\##1{{\bf #1}}
\def\=#1{\underline{\underline #1}}

\def\eps{\epsilon}
\def\epso{\epsilon_0}
\def\muo{\mu_0}
\def\ko{k_0}
\def\lambdao{\lambda_0}
\def\etao{\eta_0}
\def\.{\mbox{ \tiny{$^\bullet$} }}

\def\epsa{\epsilon_a}
\def\epsb{\epsilon_b}

\def\ux{{\#u}_x}
\def\uy{{\#u}_y}
\def\uz{{\#u}_z}

\def\un{{\#u}_n}
\def\ut{{\#u}_\tau}
\def\ub{{\#u}_b}

\def\le{\left(}
\def\ri{\right)}
\def\les{\left[}
\def\ris{\right]}
\def\lec{\left\{}
\def\ric{\right\}}

\def\c#1{\cite{#1}}
\def\l#1{\label{#1}}
\def\r#1{(\ref{#1})}

\begin{document}

\begin{center}

{\bf {\LARGE  Empirical model of optical sensing via spectral shift
of circular Bragg phenomenon}}

\vspace{5mm}

{\bf {\large Tom G. Mackay${}^{a,b,}$\footnote{Email:
T.Mackay@ed.ac.uk} and Akhlesh
Lakhtakia${}^{b,c,}$\footnote{Email: akhlesh@psu.edu}}}\\

\vspace{5mm}

${}^{a}$School of Mathematics and
   Maxwell Institute for Mathematical Sciences,\\
University of Edinburgh, Edinburgh EH9 3JZ, UK \\
${}^{b}$NanoMM---Nanoengineered Metamaterials Group, Department of Engineering Science and Mechanics,\\
Pennsylvania State University, University Park, PA 16802-6812, USA\\
${}^{c}$Department of Physics,  Indian Institute of Technology
Kanpur, Kanpur 208016, India

\end{center}
\vspace{5mm}

\begin{abstract}
Setting up an empirical model of optical sensing to exploit the circular
Bragg phenomenon displayed by chiral
sculptured thin films (CSTFs), we considered a CSTF with and without a
central twist defect of $\pi/2$ radians. The circular Bragg
phenomenon of the defect-free CSTF, and the spectral hole in the
co-polarized reflectance spectrum of the CSTF with the twist defect, were both
found to be acutely sensitive to the refractive index of  a fluid
which infiltrates the void regions of the CSTF. These findings bode
well for the  deployment of CSTFs as optical sensors.

\end{abstract}

\section{Introduction} \l{intro}

By means of physical vapor deposition, an array of parallel helical
nanowires---known as a   chiral sculptured thin film
(CSTF)---may be grown on a substrate \c{LMBR,HW}.
 At optical wavelengths, a CSTF may be regarded as a
unidirectionally nonhomogeneous continuum. CSTFs display
orthorhombic symmetry locally whereas they are  structurally chiral
from a global perspective \cite[Chap. 9]{STF_Book}. There are two
key attributes which render CSTFs attractive for a host of applications.
Firstly, CSTFs  exhibit the circular Bragg phenomenon (just as
cholesteric liquid crystals do \c{Gennes}). Thus, a structurally
right/left-handed CSTF of sufficient thickness almost completely
reflects right/left-circularly polarized (RCP/LCP) light which is
normally incident,
 but normally  incident LCP/RCP light is reflected very little, when the
 free-space wavelength lies within the Bragg regime.
 This property has led to the use of  CSTFs as circular
polarization  \c{Wu}, spectral-hole \c{Hodgkinson00}, and \u{S}olc
 \c{Ertekin} filters,  among other applications \c{Polo,LDHX}. Secondly,
CSTFs are porous, and their multiscale porosity can be
tailored to allow only species of certain shapes and sizes to
infiltrate their void regions \c{Messier}. This engineered porosity,
combined with the circular Bragg phenomenon, makes CSTFs attractive
as platforms for light-emitting devices
 with precise control over the circular polarization
state and the emission wavelengths \c{XLLCH,Zhang}, and optical
biosensors \c{L01,ML_CSTF}.

Further possibilities for a CSTF emerge if a structural twist defect is
introduced. For example, if the upper half of a CSTF is twisted by
$\pi/2$ radians about the axis of nonhomogeneity relative to the lower half, then
the co-polarized reflectance spectrum contains a spectral hole in the middle
of the Bragg regime
\c{Yang_PRE,LM99}. This phenomenon may be exploited for
narrow-bandpass filtering \c{Hod00}, as well as for optical sensing of
fluids which infiltrate the void regions of the CSTF \c{LMSWH01,Horn}.

We devised an empirical model yielding the sensitivity of
a CSTF's optical response---as a circular Bragg filter---to the refractive index of a fluid which
infiltrates the CSTF's void regions, with a view to optical-sensing
applications. We considered both a defect-free CSTF and a CSTF with a central $\pi/2$-twist defect.
Our model is not limited to normal incidence but
also encompasses oblique incidence, and we present computed results
for slightly off-normal incidence,
a realistic situation for sensing applications. Furthermore, because the material that is deposited
as the CSTF is not precisely the same as the bulk material that is
evaporated, we use an inverse homogenization procedure \cite{ML_inverse_homog}
on related but uninfiltrated columnar thin
films (CTFs) \cite{HWH_AO} to predict the spectral shifts due to infiltration of CSTFs.

An $\exp(-i\omega t)$ time-dependence is implicit, with $\omega$
denoting the angular frequency and $i = \sqrt{-1}$. The free-space
wavenumber, the free-space wavelength, and the intrinsic impedance
of free space are denoted by $\ko=\omega\sqrt{\epso\muo}$,
$\lambdao=2\pi/\ko$, and $\etao=\sqrt{\muo/\epso}$, respectively,
with $\muo$ and $\epso$ being  the permeability and permittivity of
free space. Vectors are in boldface, dyadics are underlined twice,
column vectors are in boldface and enclosed within square brackets,
and
matrixes are underlined twice and square-bracketed. %The asterisk denotes the complex conjugate, and
The Cartesian unit vectors are identified as $\ux$, $\uy$, and
$\uz$.

\section{Empirical Model}

\subsection{Uninfiltrated defect-free CSTF}
Let us begin the explication of our empirical model with a defect-free
CSTF with vacuous void regions; i.e., an uninfiltrated CSTF. The $z$
direction is taken to be the direction of nonhomogeneity. The CSTF
is supposed to have been grown on a planar substrate through the deposition of an
evaporated bulk material \cite{STF_Book}. The substrate, which lies  parallel to the
plane $z=0$, is supposed to have been rotated about the $z$ axis at a uniform angular
speed throughout the deposition process. The rise angle of each
resulting helical nanowire, relative to the $xy$ plane,  is denoted
by $\chi$. The refractive index of the deposited
material~---~assumed to be an isotropic dielectric material~---~is
written as $n_s$, which can be different from the refractive index
of the bulk material that was evaporated \c{MTR1976,BMYVM,WRL03}.

Each nanowire of a CSTF  can be modeled as a string of highly elongated
ellipsoidal inclusions, wound end-to-end around the $z$ axis to
create a helix \cite{Sherwin,Lakh_Opt}.  The surface of each
ellipsoidal inclusion is characterized by the  shape dyadic
\begin{equation}
 \un \, \un + \gamma_\tau \, \ut \, \ut + \gamma_b \, \ub \,
\ub ,
\end{equation}
wherein the normal, tangential and binormal basis vectors are given
as
\begin{equation}
\left. \begin{array}{l}
 \un = - \ux \, \sin \chi + \uz \, \cos \chi \vspace{4pt} \\
 \ut =  \ux \, \cos \chi + \uz \, \sin \chi \vspace{4pt} \\
\ub = - \uy
\end{array}
\right\}.
\end{equation}
By choosing the shape parameters $\gamma_{b} \gtrsim 1$ and
$\gamma_\tau \gg 1$, an aciculate shape is imposed on the inclusions. For the
numerical results presented in \S\ref{Numerica}, we fixed $\gamma_\tau =
15$ while noting that  increasing $\gamma_\tau$ beyond 10 does not
give rise to significant effects for slender inclusions
\cite{Lakh_Opt}. The helical nanowires occupy only a proportion  $f
\in \le 0, 1 \ri $ of the total CSTF volume;   the volume fraction
of the CSTF not occupied by nanowires is $1 - f$.

 At length scales
much greater than the nanoscale, the CSTF's relative permittivity
dyadic may be expressed as
\begin{equation}
\=\eps_{\,1} = {\=S}_{\,z} \le h  \frac{\pi z}{\Omega} \ri \.
{\=S}_{\,y} \le \chi \ri \. \=\eps^{ref}_{\,1} \.
{\=S}^T_{\,y} \le \chi \ri \. {\=S}^T_{\,z} \le h \frac{\pi
z}{\Omega} \ri, \l{eps1_dyadic}
\end{equation}
where $2 \Omega$ is the structural period and the rotation dyadics
\begin{equation}
\left.
\begin{array}{l}
{\=S}_{\,y} \le \chi \ri = \#u_y\, \#u_y + \le \#u_x\, \#u_x +
\#u_z\, \#u_z \ri \cos \chi + \le \#u_z\, \#u_x - \#u_x\, \#u_z \ri
\sin \chi \vspace{4pt} \\
{\=S}_{\,z} \le \sigma \ri =
 \#u_z\, \#u_z +
\le \#u_x\, \#u_x + \#u_y\, \#u_y \ri \cos  \sigma  + \le \#u_y\,
\#u_x - \#u_x\, \#u_y \ri \sin  \sigma
\end{array}
\right\}.
\end{equation}
The handedness parameter $h = + 1$ for a structurally right-handed
CSTF, and $h = - 1$ for a structurally left-handed CSTF.
 The reference
relative permittivity dyadic $\=\eps^{ref}_{\,1}$ has the
orthorhombic form
\begin{equation}
\=\eps^{ref}_{\,1} = \eps_{a1}   \,\un\,\un +\eps_{b1}\,\ut\,\ut \,
+\,\eps_{c1}\,\ub\,\ub . \l{eps1_ref_dyadic}
\end{equation}

The nanowire rise angle $\chi$ can be measured from
scanning-electron-microscope imagery. In principle, the relative
permittivity parameters $\lec \eps_{a1}, \eps_{b1}, \eps_{c1} \ric$
of an uninfiltrated CSTF are also measurable. However, in view of
the paucity of suitable experimental data on CSTFs, our empirical
model relies on the measured experimental data on the related
columnar thin films (CTFs). In order to deposit both CSTFs and CTFs,
the vapor flux is directed at a fixed angle $\chi_v$ with respect to the
substrate plane. The different morphologies of CSTFs and CTFs are due
to the rotation of the substrate for the former but not for the
latter. The parameters $\lec \eps_{a1}, \eps_{b1}, \eps_{c1}, \chi
\ric$ are functions of $\chi_v$.

The nanoscale
model parameters $\lec n_s, f, \gamma_b \ric$ are not readily
determined  by experimental means. However,  the process of inverse
homogenization can be employed to determine these parameters from a
knowledge of $\lec \eps_{a1}, \eps_{b1}, \eps_{c1} \ric$, as was done for
titanium-oxide  CTFs
in a
predecessor paper \c{ML_inverse_homog}.

\subsection{Infiltrated defect-free CSTF}

With optical-sensing applications in mind, next we consider the effect of
filling the void regions of a defect-free CSTF
 with a fluid of
refractive index $n_\ell$. This brings about a change in the
reference relativity permittivity dyadic. The infiltrated CSTF is
characterized by the  relative permittivity dyadic $\=\eps_{\,2}$,
which has the same eigenvectors as $\=\eps_{\,1}$ but  different
eigenvalues. Thus, the infiltrated CSTF is characterized by
\r{eps1_dyadic} and  \r{eps1_ref_dyadic}, but with $\=\eps_{\,2}$ in
lieu of $\=\eps_{\,1}$, $\=\eps^{ref}_{\,2}$ in lieu of
$\=\eps^{ref}_{\,1}$ and $\lec \eps_{a2}, \eps_{b2}, \eps_{c2} \ric$
in lieu of $\lec \eps_{a1}, \eps_{b1}, \eps_{c1} \ric$. The nanowire
rise angle $\chi$ remains unchanged.

In our model, the Bruggeman homogenization formalism is applied in its usual
forward sense \c{EAB} to determine $\lec \eps_{a2}, \eps_{b2},
\eps_{c2} \ric$, from knowledge of the nanoscale model parameters
$\lec n_s, f, \gamma_b \ric$ together with $\lec
n_\ell,\gamma_\tau\ric$, as described elsewhere
\c{Lakh_Opt}.

\subsection{Boundary-value problem} \l{bvp}

Let us now suppose that a CSTF  occupies the region $0 \leq z \leq
L$, with the half-spaces $z< 0$ and $z > L$ being vacuous. An
arbitrarily polarized plane wave is incident on the CSTF from the
half-space $z < 0$. Its wavevector lies in the $xz$ plane, making an
angle $\theta \in \les 0, \pi/2 \ri$ relative to the $+z$ axis. As a
result, there is a reflected plane wave in the half-space $z < 0$
and a transmitted plane wave in the half-space $z > L$. Thus, the
total electric field phasor in the half-space $z < 0$ may be
expressed as
\begin{eqnarray}
\nonumber
 \#E (\#r) &=& \les a_L \frac{i \#u_y - \#p_+}{\sqrt{2}} - a_R
\frac{i \#u_y + \#p_+}{\sqrt{2}} \ris \,\exp\le{i\kappa}x\ri\, \exp \le i \ko z\cos\theta \ri
\\
&&\quad- \les
r_L \frac{i \#u_y - \#p_-}{\sqrt{2}} - r_R \frac{i \#u_y +
\#p_-}{\sqrt{2}} \ris \, \exp\le{i\kappa}x\ri\, \exp \le - i \ko z \cos\theta\ri,  \quad
 z < 0,
\end{eqnarray}
while that in the half-space $z > L$ may be expressed as
\begin{equation}
\#E (\#r) = \les t_L \frac{i \#u_y - \#p_+}{\sqrt{2}} - t_R \frac{i
\#u_y + \#p_+}{\sqrt{2}} \ris \, \exp\le{i\kappa}x\ri\, \exp \les i \ko \le z - L \ri \cos\theta\ris,
 \quad z > L,
\end{equation}
wherein $\#p_\pm = \mp \#u_x \cos \theta + \#u_z \sin \theta$ and $\kappa=\ko\sin\theta$.

Our aim is to determine the unknown amplitudes $r_L$ and $r_R$ of
the LCP and RCP components of the reflected plane wave, and the
unknown amplitudes $t_L$ and $t_R$ of the LCP and RCP components of
the transmitted plane wave, from the known amplitudes $a_L$ and
$a_R$ of the LCP and RCP components of the incident plane wave. As
is comprehensively described elsewhere \c{STF_Book}, this is
achieved by solving the 4$\times$4 matrix/4-vector relation
\begin{equation} \l{main_eq}
 [\#f^{exit}]  =
[\=M(L)]  \. [\#f^{entry}].
\end{equation}
Here, the column 4-vectors
\begin{equation}
[\#f^{entry}] = \frac{1}{\sqrt{2}} \le
\begin{array}{c}
\le r_L + rR \ri + \le a_L + a_R \ri \vspace{4pt} \\
i \les - \le r_L - rR \ri + \le a_L - a_R \ri \ris \vspace{4pt} \\
-i \les  \le r_L - rR \ri + \le a_L - a_R \ri \ris/ \etao \vspace{4pt} \\
- \les \le r_L + rR \ri - \le a_L + a_R \ri \ris/ \etao
\end{array}
 \ri, \quad
 [\#f^{exit}]  = \frac{1}{\sqrt{2}} \le
\begin{array}{c}
 t_L + tR  \vspace{4pt} \\
i  \le t_L - tR \ri \vspace{4pt} \\
-i   \le t_L - tR \ri / \etao \vspace{4pt} \\
  \le t_L + tR \ri / \etao
\end{array}
 \ri,
\end{equation}
arise from the field phasors at $z=0$ and $z=L$, respectively. The
optical response characteristics of the CSTF are  encapsulated by
the 4$\times$4 transfer matrix $[\=M(L)] $, which is conveniently expressed as \c{LVM}
\begin{equation} \l{m_eq}
[\=M(L)]  =
[\=B(h \frac{\pi z}{\Omega})]
\. [\=M'(L)] ,
\end{equation}
wherein the 4$\times$4 matrix
\begin{equation}
[\=B(\sigma)] = \le
\begin{array}{cccc}
\cos  \sigma  & - \sin  \sigma  & 0 & 0 \\  \sin
 \sigma  & \cos  \sigma  &0 & 0 \\
0 & 0& \cos  \sigma  & - \sin  \sigma  \\
0 & 0 &  \sin  \sigma  & \cos  \sigma
\end{array}
 \ri .
\end{equation}
The 4$\times$4 matrizant $ [\=M'(z)] $ satisfies the ordinary
differential equation
\begin{equation}
\frac{d}{dz}  [\=M'(z)]  = i \,
 [\=P'(z)]  \.
 [\=M'(z)] ,  \l{MODE}
\end{equation}
subject to the boundary condition $ [\=M'(0)]  =
[\=I]$,
with $[\=I]$ being the identity 4$\times$4 matrix. The 4$\times$4
matrix \c{LVM}
\begin{equation}
\nonumber
 [\=P'(z)]=
\begin{bmatrix}
0 & -i h\frac{\pi}{\Omega} & 0 & \omega\muo\\
i h \frac{\pi}{\Omega} & 0 & -\omega\muo & 0\\
0 & -\omega\epso\epsilon_{\rm c}  & 0 & - i h \frac{\pi}{\Omega}\\
\frac{\omega\epso\epsb}{\tau} & 0 & i h \frac{\pi}{\Omega} & 0
\end{bmatrix}
\, +
\end{equation}
\begin{equation}
\l{kermat}
\begin{bmatrix}
-\frac{\kappa \le \epsilon_{b} - \epsilon_{a} \ri }{2\epsilon_{a}
\tau} \cos\xi \sin{2\chi} & 0 &
-\frac{\kappa^2}{\omega\epso\epsilon_{a}\tau}\sin\xi\cos\xi &
-\frac{\kappa^2}{\omega\epso\epsilon_{a}\tau}\cos^2\xi \\
 & & & \\
\frac{\kappa \le \epsilon_{b}-\epsilon_{a}
\ri } {2\epsilon_{a} \tau}\sin\xi\sin{2\chi} & 0 &
\frac{\kappa^2}{\omega\epso\epsilon_{a}\tau}\sin^2\xi &
\frac{\kappa^2}{\omega\epso\epsilon_{a}\tau}\sin\xi\cos\xi \\
 & & & \\
 \frac{\kappa^2}{\omega\muo}\sin\xi\cos\xi &
 \frac{\kappa^2}{\omega\muo}\cos^2\xi & 0 & 0 \\
 & & & \\
 - \frac{\kappa^2}{\omega\muo}\sin^2\xi &
-\frac{\kappa^2}{\omega\muo}\sin\xi\cos\xi &
-\frac{\kappa\le \epsilon_{b} - \epsilon_{a}
\ri}{2\epsilon_{a}\tau}\sin\xi\sin{2\chi} & -\frac{\kappa\le
\epsilon_{b} - \epsilon_{a}
\ri}{2\epsilon_{a}\tau}\cos\xi\sin{2\chi}
\end{bmatrix}
\,
\end{equation}
containing
\begin{equation}
\xi =  h \pi z/\Omega  ,\qquad
\tau =  \cos^2\chi+ \le \epsb/\epsa \ri \sin^2\chi
\end{equation}
depends on whether the CSTF is uninfiltrated or infiltrated.
Equation~\r{MODE} can
be solved for $[\=M'(z)]$ by numerical
means, most conveniently using a piecewise uniform approximation
\c{STF_Book}.

Once $[\=M'(z)]$ is determined,  it is
a straightforward matter of linear algebra to extract the reflection
amplitudes $r_{L,R}$ and transmission amplitudes $t_{L,R}$ from
\r{main_eq}, for specified incident amplitudes $a_{L,R}$. Following
the standard convention, we introduce the reflection coefficients
$r_{LL,LR,RL,RR}$ and transmission coefficients $t_{LL,LR,RL,RR}$
per
\begin{equation}
\le \begin{array}{c} r_L \vspace{4pt} \\
r_R \end{array} \ri = \le \begin{array}{cc} r_{LL}  & r_{LR}
\vspace{4pt} \\ r_{RL} & r_{RR}
\end{array} \ri
 \le \begin{array}{c} a_L \vspace{4pt} \\
a_R \end{array} \ri,  \qquad \quad
\le \begin{array}{c} t_L \vspace{4pt} \\
t_R \end{array} \ri = \le \begin{array}{cc} t_{LL}  & t_{LR}
\vspace{4pt} \\ t_{RL} & t_{RR} \end{array} \ri
\le \begin{array}{c} a_L \vspace{4pt} \\
a_R \end{array} \ri.
\end{equation}
The square magnitude of a reflection or transmission coefficient
yields the corresponding reflectance or transmittance; i.e.,
$R_{\alpha \beta} = \left| r_{\alpha \beta} \right|^2$ and $T_{\alpha
\beta} = \left| t_{\alpha \beta} \right|^2$, where $\alpha, \beta \in
\lec L, R \ric$.

\subsection{CSTF with central $\pi/2$-twist defect} \l{twisted_CSTF}

As mentioned in \S\ref{intro}, the introduction of a central
twist defect leads to a narrowband feature that can be very useful for
sensing applications.
 Therefore, we further  consider the CSTF of finite thickness introduced in
\S\ref{bvp}  but here with the upper half $z \in \les L/2, L \ris$
of the CSTF twisted
 about the $z$ axis by an angle of $\pi/2$ radians with respect to the lower half $z \in \les 0, L/2 \ri$.

Mathematically,  the central $\pi/2$-twist defect is accommodated as follows: The
relative permittivity dyadic of the centrally twisted uninfiltrated
CSTF is per \r{eps1_dyadic} but with the rotation dyadic $
{\=S}_{\,z} \le h  \frac{\pi z}{\Omega} \ri $ therein replaced
by $ {\=S}_{\,z} \lec h \les \frac{\pi z}{\Omega} + \psi(z) \ris
\ric $, where
\begin{equation}
\psi(z) = \left\{ \begin{array}{lcr} 0, && 0 \leq z < L/2
\vspace{6pt}
\\ \pi/2, && L/2 \leq z \leq L \end{array}
\right..
\end{equation}
 The relative permittivity dyadic of the centrally twisted and
infiltrated CSTF follows from the corresponding dyadic for the
centrally twisted and uninfiltrated CSTF, in exactly the same way as is
the case when the CSTF  is defect-free. The calculation of the
reflectances and transmittances follows the same path as is
described in \S\ref{bvp} with the exception that \r{m_eq} therein is
replaced by
\begin{equation} \l{new_m_eq}
[\=M(L)]
 =
[\=B \le  h \frac{\pi L
}{2 \Omega}+\frac{\pi}{2} \ri]\. [\=M'(L/2)]
\. [\=B \le h \frac{\pi L
}{2 \Omega}-\frac{\pi}{2} \ri]\.  [\=M'(L/2)] .
\end{equation}

\section{Numerical results} \l{Numerica}

In order to illustrate the empirical model, we chose a CSTF of
thickness $L = 40 \Omega$ where the structural half-period $\Omega =
185$ nm.   The chosen relative permittivity parameters, namely
\begin{equation}
\left.
\begin{array}{l}
\eps_{a1} = \displaystyle{\les 1.0443 + 2.7394 \le \frac{2
\chi_v}{\pi} \ri - 1.3697
\le \frac{2 \chi_v}{\pi} \ri^2 \ris^2} \vspace{6pt} \\
\eps_{b1} = \displaystyle{ \les 1.6765 + 1.5649 \le \frac{2
\chi_v}{\pi} \ri - 0.7825 \le \frac{2 \chi_v}{\pi} \ri^2 \ris^2}
 \vspace{6pt} \\
\eps_{c1} = \displaystyle{ \les 1.3586 + 2.1109 \le \frac{2
\chi_v}{\pi} \ri - 1.0554 \le \frac{2 \chi_v}{\pi} \ri^2 \ris^2}
\end{array}
\right\} \l{tio1}
\end{equation}
with
\begin{equation}
\chi = \tan^{-1} \le {2.8818}   \tan \chi_v \ri \l{tio2},
\end{equation}
emerged from data measured for a CTF made by evaporating
patinal${}^{\mbox{\textregistered}}$ titanium oxide \c{HWH_AO}.
These relations---which came from measurements at $\lambdao=
633$~nm---were presumed to be constant over the range of wavelengths
considered here. Values for the corresponding nanoscale model
parameters $\lec n_s, f, \gamma_b \ric$,  as computed  using the
inverse Bruggeman homogenization formalism \c{ML_inverse_homog}, are
provided in Table~1 for the vapor flux angles  $\chi_v = 15^\circ$,
$30^\circ$, and $60^\circ$. Furthermore, we set $h=+1$. The angle of
incidence $\theta$ was fixed at $10^\circ$.

\begin{figure}[!h]
\centering
\includegraphics[width=2.6in]{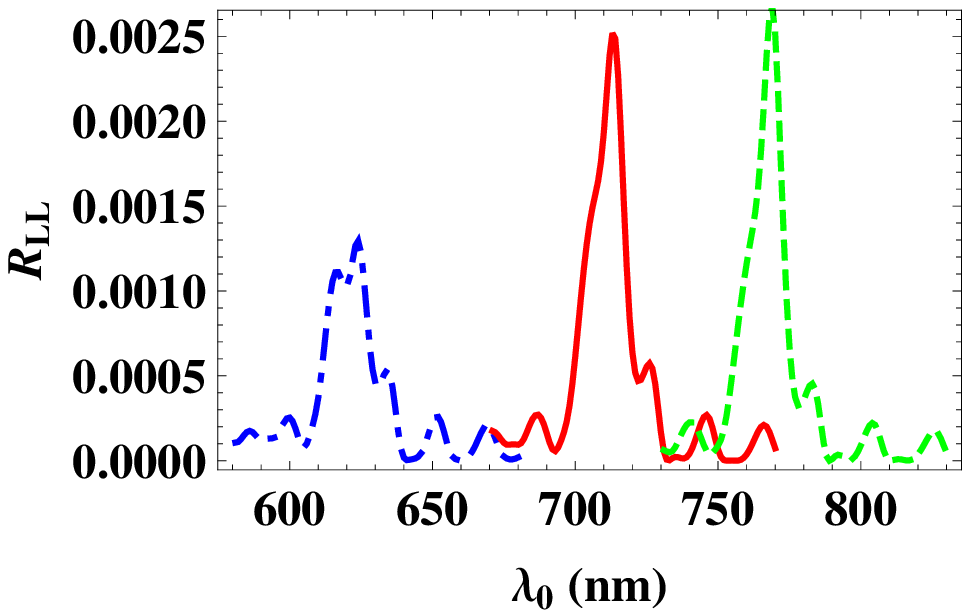} \hfill
\includegraphics[width=2.6in]{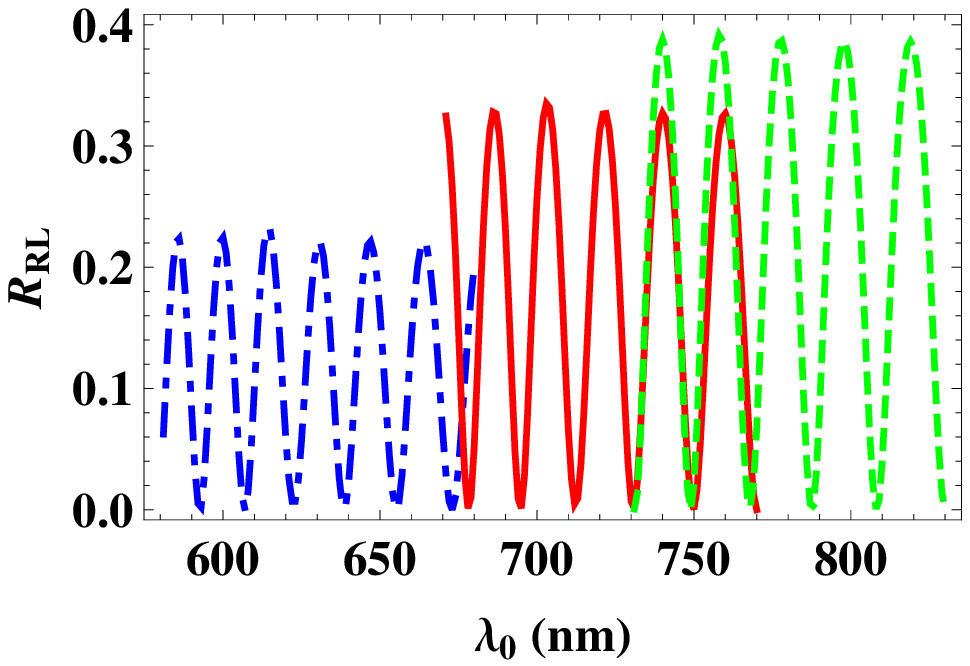}\\
\includegraphics[width=2.6in]{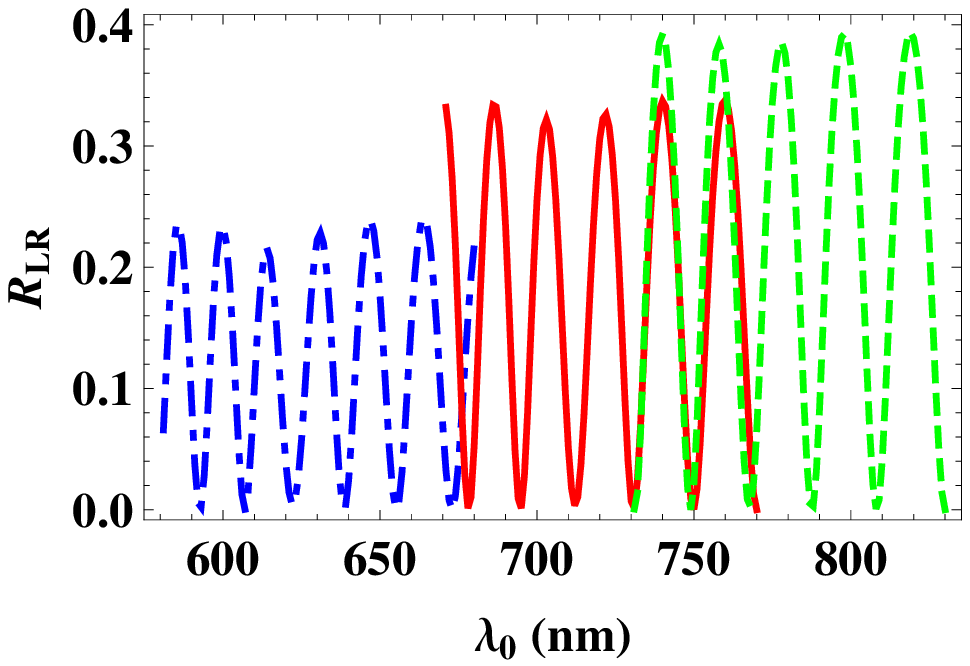} \hfill
\includegraphics[width=2.6in]{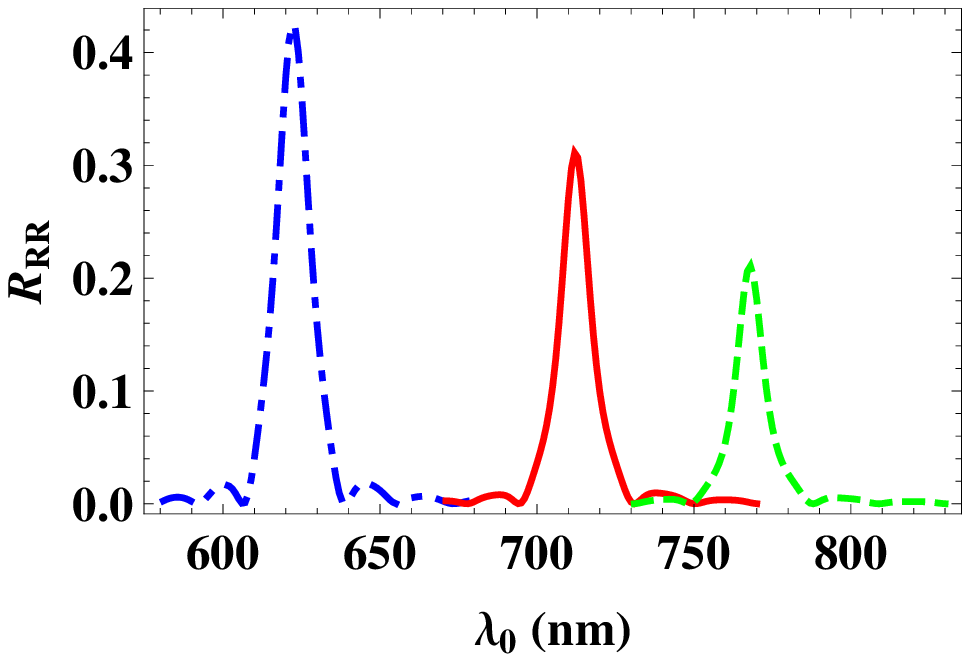}\\
\includegraphics[width=2.6in]{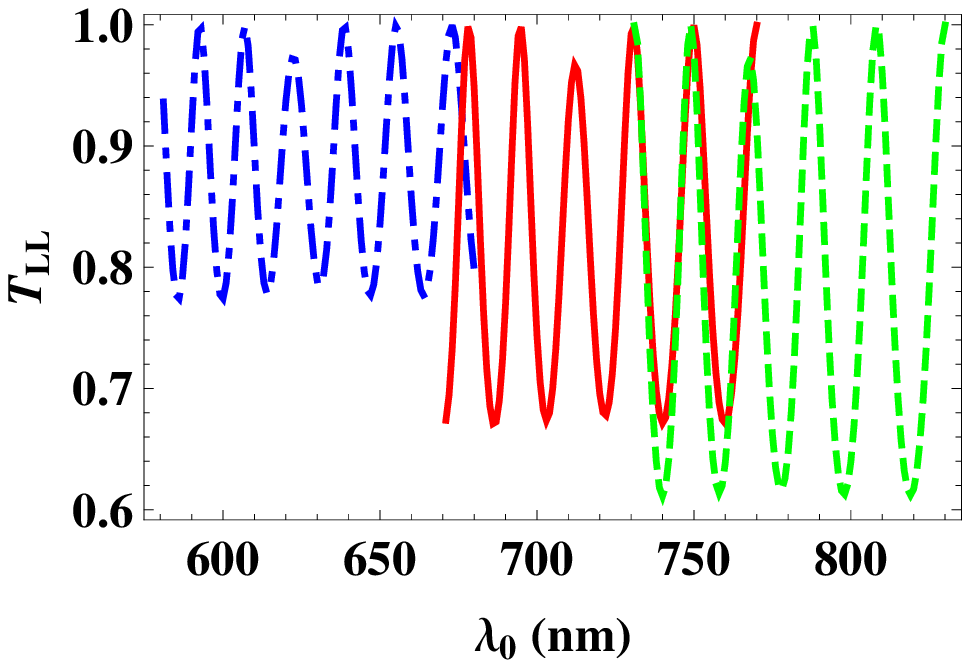} \hfill
\includegraphics[width=2.6in]{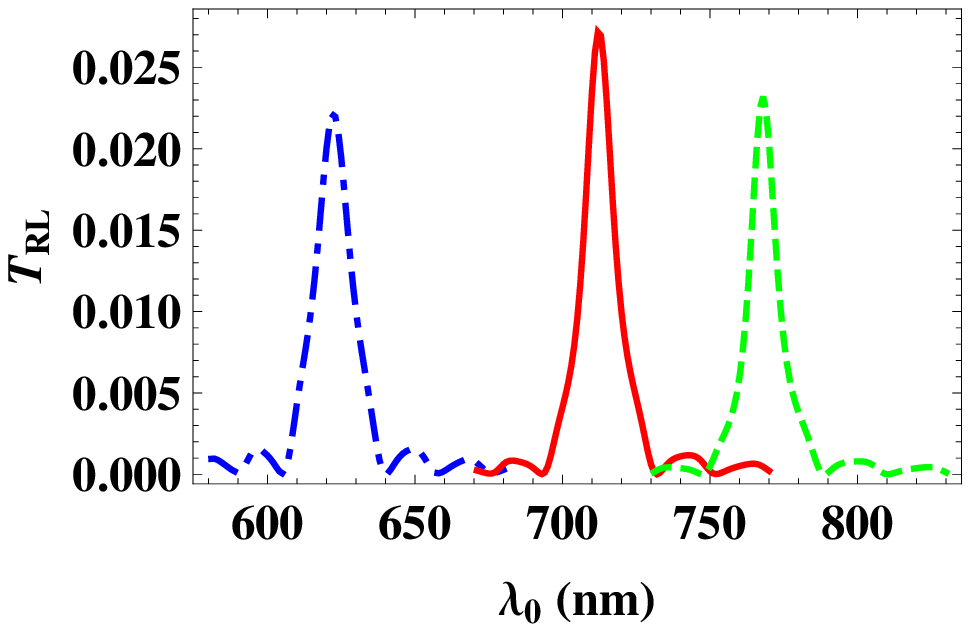}\\
\includegraphics[width=2.6in]{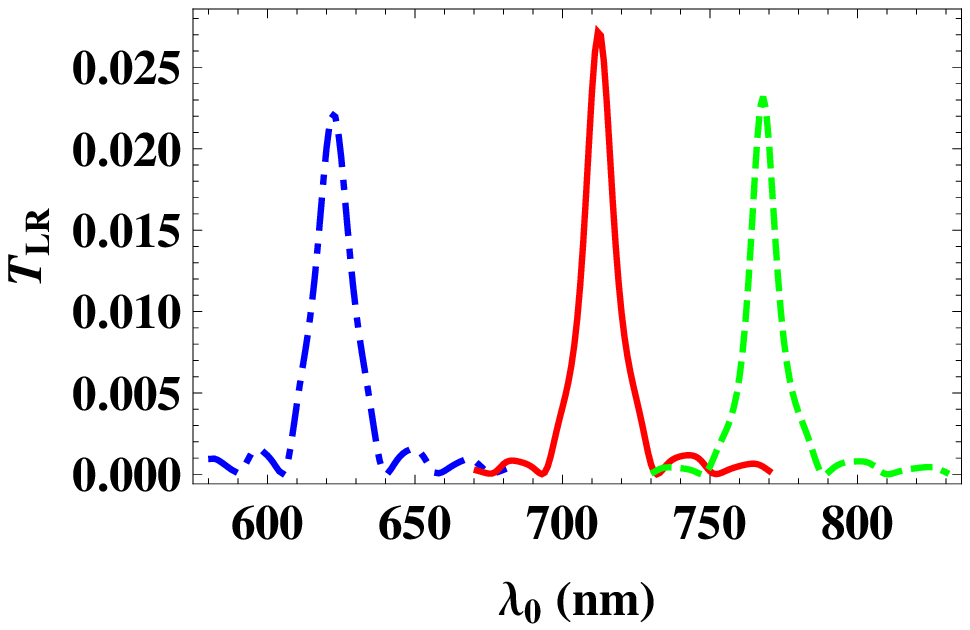} \hfill
\includegraphics[width=2.6in]{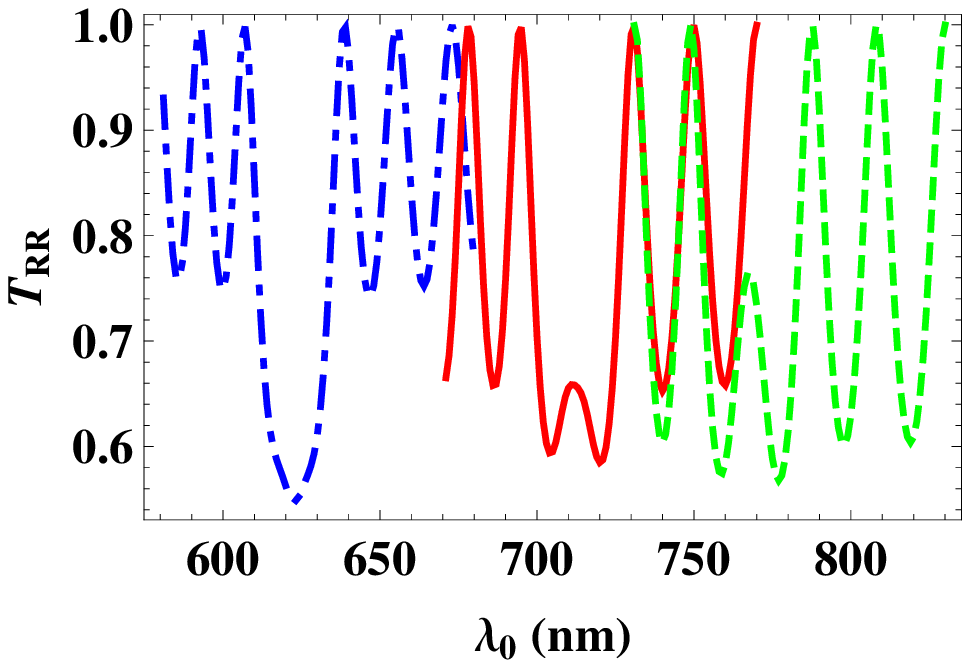}
\caption{Reflectances and transmittances  plotted against the
free-space wavelength for a defect-free titanium-oxide CSTF; $L = 40
\Omega$, $ \Omega = 185$ nm,   $h=+1$, $\chi_v = 15^\circ$, and
$\theta = 10^\circ$. The CSTF is infiltrated with a fluid of
refractive index $n_\ell = 1.0$ (blue broken-dashed curves), $1.3$
(red solid curves), and 1.5 (green dashed curves). } \label{Fig1}
\end{figure}

Computed reflectances and transmittances are plotted versus $\lambdao$
in Fig.~\ref{Fig1}, for the defect-free CSTF for which we set
$ \chi_v =
 15^\circ$. Further computations (not presented here) using
 other values of $\chi_v$  revealed
  qualitatively similar graphs of reflectances and transmittances versus $\lambdao$.
   The effects of three
  values of $n_\ell$---namely,
  $n_\ell = 1$, $1.3$ and $1.5$---are represented in Fig.~\ref{Fig1}. The circular Bragg phenomenon is
  most obviously appreciated as a sharp local maximum in the graphs of
   $R_{RR}$, with attendant features occurring in the graphs of some other reflectances and
  transmittances.
If $\lambda^{max}_0 $ denotes the free-space wavelength corresponding
to this local
 maximum,  from Fig.~\ref{Fig1}, we found that  $\lambdao^{max} \approx 622$
  nm for $n_\ell = 1$, $\lambdao^{max} \approx 712$
  nm for $n_\ell = 1.3$, and $\lambdao^{max} \approx 768$
  nm for $n_\ell = 1.5$.

Clearly, the circular Bragg
 phenomenon undergoes a substantial spectral shift  as $n_\ell$ increases from unity. In order to
 elucidate
  further this matter, we focused on the
spectral-shift sensitivity $d \lambda^{max}_0 / d
 n_\ell$.
Graphs of  $d \lambda^{max}_0 / d
 n_\ell$
against $\lambda^{max}_0$,  computed
  for the  range $1 < n_\ell < 1.5
 $,  are presented in Fig.~\ref{Fig1a}. In addition to results for the
vapor flux angle $\chi_v =
15^\circ$, results are also plotted in Fig.~\ref{Fig1a} for   $\chi_v =
 30^\circ$ and  $ \chi_v =
 60^\circ$.
For all vapor flux angles considered and all values of $n_\ell \in \le
1, 1.5 \ri$, the spectral-shift sensitivity  $d \lambda^{max}_0 / d
 n_\ell$ is positive-valued and greater than 118 nm per refractive index unit
 (RIU).
When $\chi_v = 15^\circ$,  $d \lambda^{max}_0 / d
 n_\ell$ generally decreases as $ \lambda^{max}_0$ increases. A
 similar trend is exhibited for $\chi_v
= 30^\circ$, but $d \lambda^{max}_0 / d
 n_\ell$ generally increases as $ \lambda^{max}_0$ increases for  $\chi_v
 =60^\circ$.

 The  center wavelength of the circular Bragg regime has been estimated as \c{VL00}
\begin{equation} \l{estimate}
\lambdao^{Br} \approx \Omega \le \sqrt{ \eps_{c2}} +
\sqrt{\frac{\eps_{a2} \eps_{b2}}{\eps_{a2} \cos^2 \chi + \eps_{b2}
\sin^2 \chi}} \ri \sqrt{ \cos \theta}.
\end{equation}
The graphs of $d \lambda^{Br}_0 / d
 n_\ell$ versus $ \lambda^{Br}_0$, as provided in Fig.~\ref{Fig1b} for the vapor flux angles
 $\chi_v = 15^\circ$, $30^\circ$, and $60^\circ$,  are remarkably
 similar (but not identical) to the graphs of  $d \lambda^{max}_0 / d
 n_\ell$ versus $ \lambda^{max}_0$ displayed in Fig.~\ref{Fig1a}.
 Thus, the center-wavelength formula \r{estimate} can yield a
 convenient estimate of the spectral-shift sensitivity, without having to solve the
 reflection-transmission problem.

We turn now to the CSTF with a central twist defect of $\pi/2$ radians, as
described in \S\ref{twisted_CSTF}. Graphs of the reflectances and
transmittances versus $\lambdao$ for  $\chi_v =
15^\circ$ are provided in
Fig.~\ref{Fig2}. As we remarked for the defect-free CSTF, graphs (not
presented here) which are qualitatively similar to those presented
in Fig.~\ref{Fig2} were obtained when other values of the vapor flux angle $\chi_v$ were
considered. The graphs of Fig.~\ref{Fig2} are substantially
different to those of Fig.~\ref{Fig1}: the local maximums in the
graphs of   $R_{RR}$  in Fig.~\ref{Fig1} have been replaced by sharp local
  minimums in Fig.~\ref{Fig2}. These local minimums---which
  represent an ultranarrowband spectral hole---arise at  the free-space wavelengths
  $\lambda^{min}_0$ that are approximately the same as
    the corresponding local maximums $\lambda^{max}_0 $ in  Fig.~\ref{Fig1}.

The location of the spectral hole on the $\lambdao$ axis  is highly sensitive to $n_\ell$. In a similar manner to before,
we explore this matter by computing the spectral-shift sensitivity
$d \lambda^{min}_0 / d
 n_\ell$ at each value of $n_\ell$. In Fig.~\ref{Fig2a},
  $d \lambda^{min}_0 / d
 n_\ell$ is plotted
against $\lambda^{min}_0$, with the spectral-shift sensitivity
computed
  for the  range $1 < n_\ell < 1.5
 $ and with  $\chi_v = 15^\circ$, $ 30^\circ$,
and $60^\circ$. The plots of $d \lambda^{min}_0 / d
 n_\ell$ versus $\lambda^{min}_0$
  in Fig.~\ref{Fig2a} are both qualitatively and quantitatively
 similar to those of $d \lambda^{max}_0 / d
 n_\ell$ versus $\lambda^{max}_0$ in Fig.~\ref{Fig1a}. That is,
positive-valued $d \lambda^{min}_0 / d
 n_\ell$ generally decreases as $\lambda^{min}_0$ increases for
$\chi_v = 15^\circ$ and $ 30^\circ$, and generally increases as
$\lambda^{min}_0$ increases for $\chi_v= 60^\circ$. A similar correspondence exists
with Fig.~\ref{Fig1b}.

\section{Closing remarks}

Our empirical model has demonstrated that the circular Bragg
phenomenon associated with a defect-free CSTF, and the
ultranarrowband spectral hole displayed
by a CSTF with a central $\pi/2$-twist defect,  both undergo
substantially large spectral shifts due to infiltration by a fluid.
Although, owing to lack of availability
of experimental data, we did not consider wavelength-dispersion in the dielectric properties
of the material used to deposit a CSTF,
the promise of
CSTFs---with or without a structural twist---to act as
platforms for optical sensing was clearly highlighted. Experimental validation
is planned.

\section*{Acknowledgements}
TGM is supported by a  Royal Academy of Engineering/Leverhulme Trust
Senior Research Fellowship. AL thanks the Binder Endowment at Penn
State for partial financial support of his research activities.

%% \ackrule

\bibliographystyle{IEEEtran}

\newpage
\begin{table}[!h]
\begin{center}
\begin{tabular}{|c  | c | c | c|} \hline \vspace{-6pt} &&&   \\
$\chi_v$ & $\gamma_b$ & $f$ & $n_s$   \\ &&& \vspace{-6pt} \\  \hline &&& \vspace{-6pt} \\
$15^\circ$ & 2.2793 & 0.3614 & 3.2510 \\ \hline &&& \vspace{-6pt}\\
$30^\circ$ & 1.8381 & 0.5039 & 3.0517 \\ \hline &&& \vspace{-6pt} \\
$60^\circ$ & 1.4054 & 0.6956 & 2.9105 \\ \hline
\end{tabular}
\caption{Nanoscale model parameters $\gamma_b$, $f$ and $n_s$ for
$\chi_v = 15^\circ$, $30^\circ$, and $60^\circ$. } \label{tab1}
\end{center}
\end{table}

\newpage

\newpage
\begin{figure}[!h]
\centering
\includegraphics[width=3.3in]{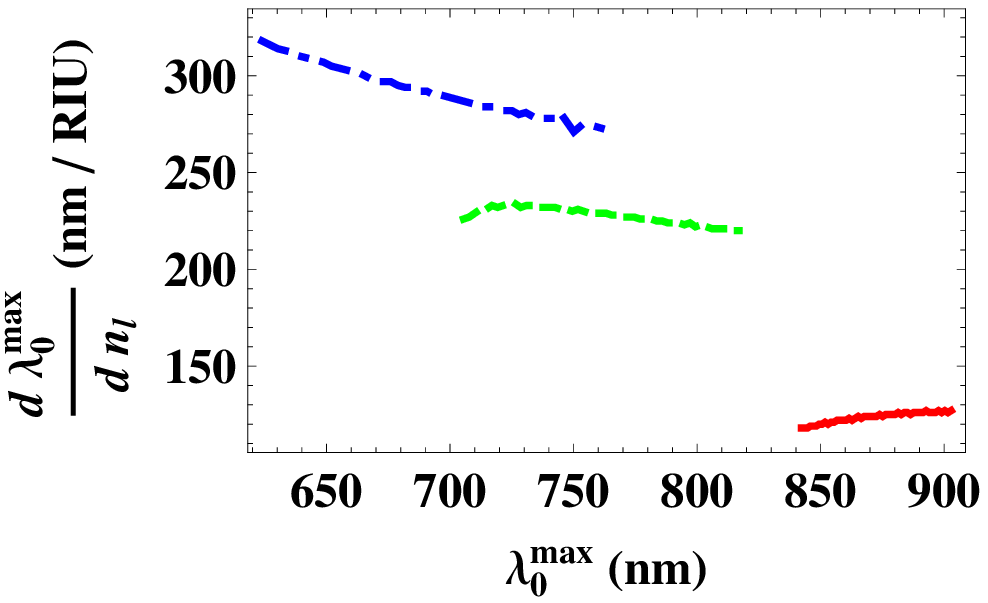}
 \caption{\l{Fig1a}
 Spectral-shift sensitivity  $d \lambda^{max}_0 / d n_\ell$
  plotted against $\lambda^{max}_0$  for $n_\ell \in \le 1, 1.5
 \ri$.
 The vapor flux angle $\chi_v =
15^\circ$ (blue broken-dashed curve), $30^\circ$ (green dashed
curve), and $60^\circ$ (red solid curve). }
\end{figure}

\newpage
\begin{figure}[!h]
\centering
\includegraphics[width=3.3in]{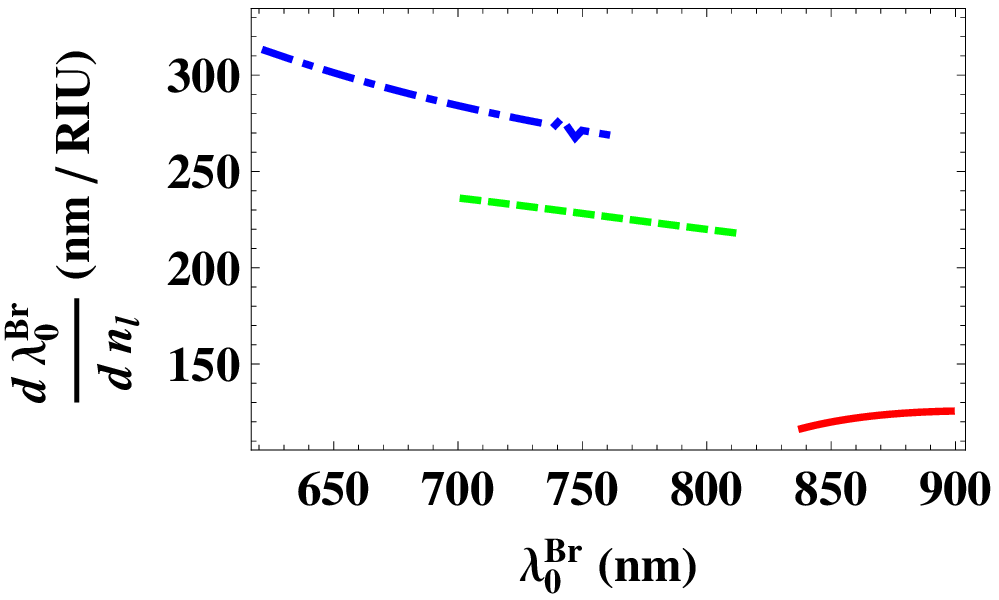}
 \caption{\l{Fig1b}
 Spectral-shift sensitivity  as estimated by $d \lambda^{Br}_0 / d n_\ell$
  plotted against $\lambda^{Br}_0$  for $n_\ell \in \le 1, 1.5
 \ri$.
 The vapor flux angle $\chi_v =
15^\circ$ (blue broken-dashed curve), $30^\circ$ (green dashed
curve), and $60^\circ$ (red solid curve). }
\end{figure}

\newpage
\begin{figure}[!h]
\centering
\includegraphics[width=2.6in]{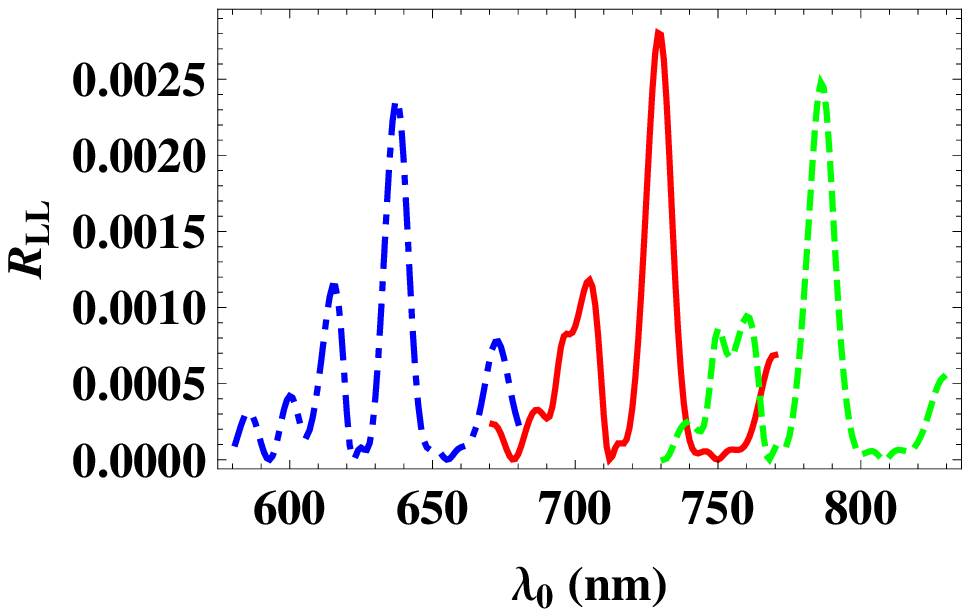} \hfill
\includegraphics[width=2.6in]{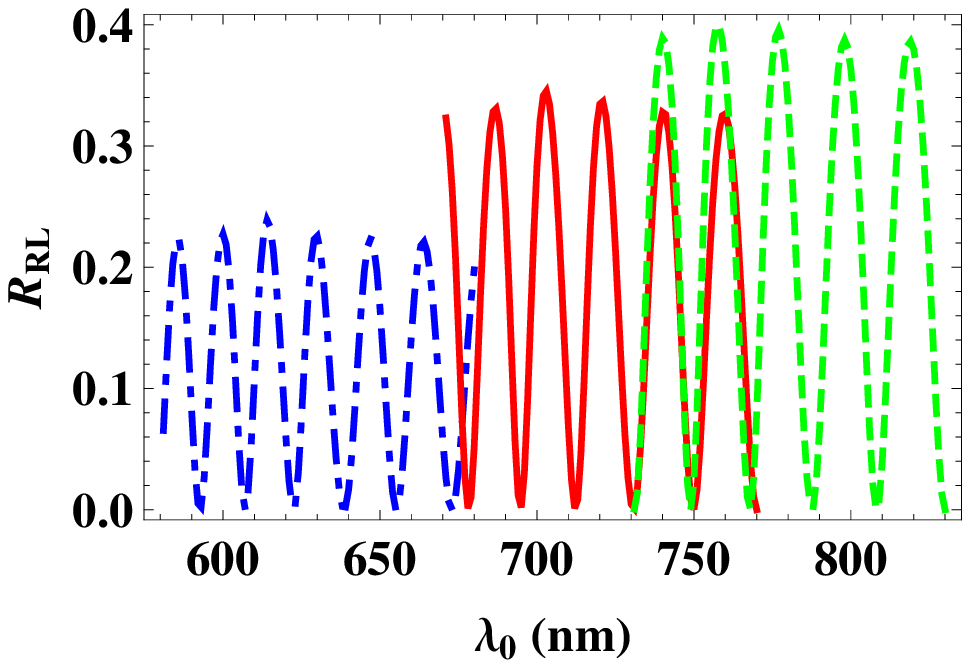}\\
\includegraphics[width=2.6in]{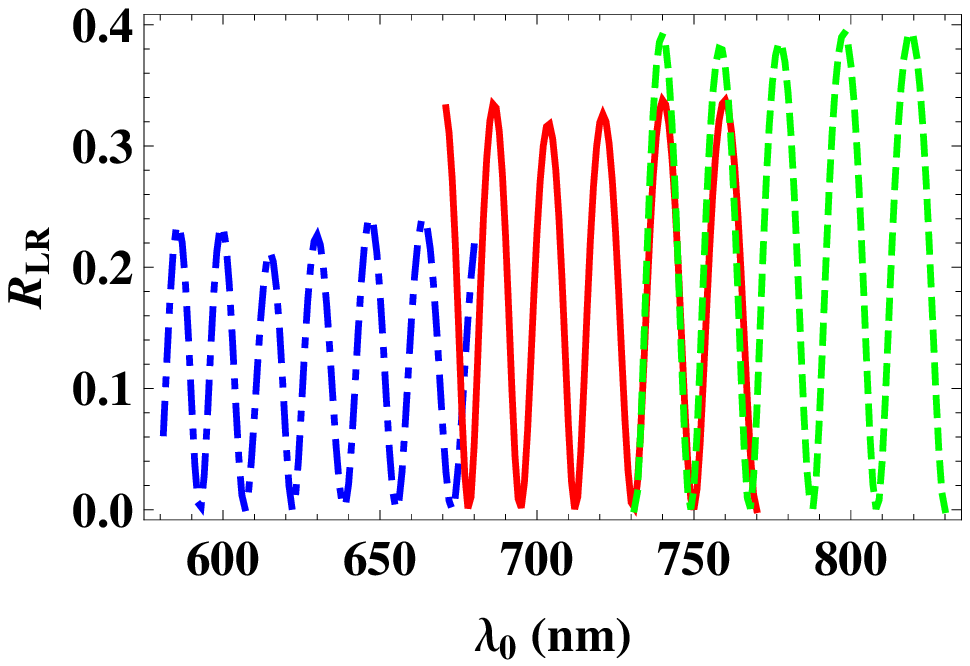} \hfill
\includegraphics[width=2.6in]{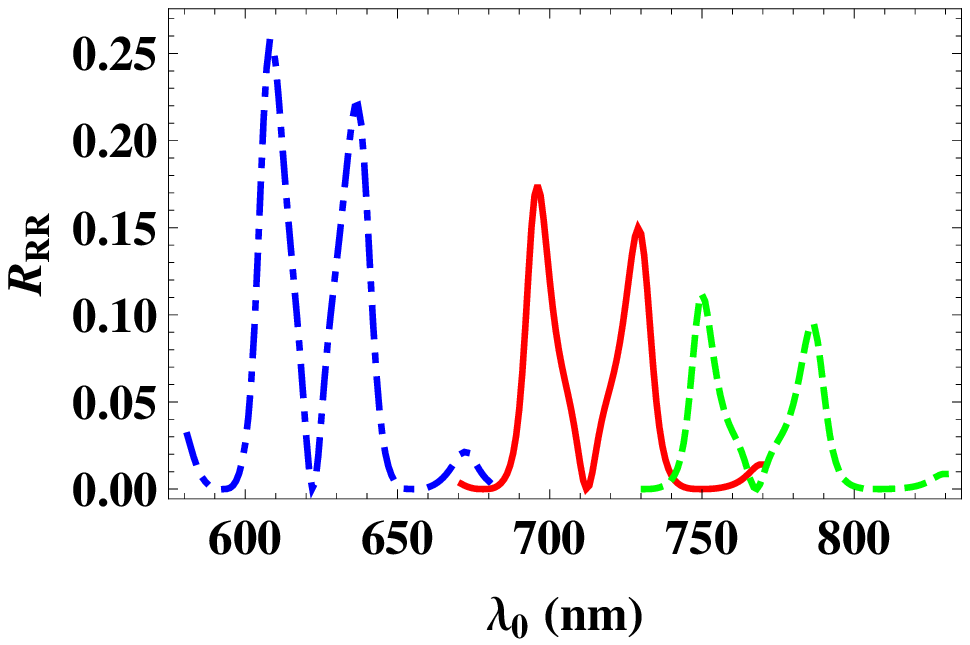}\\
\includegraphics[width=2.6in]{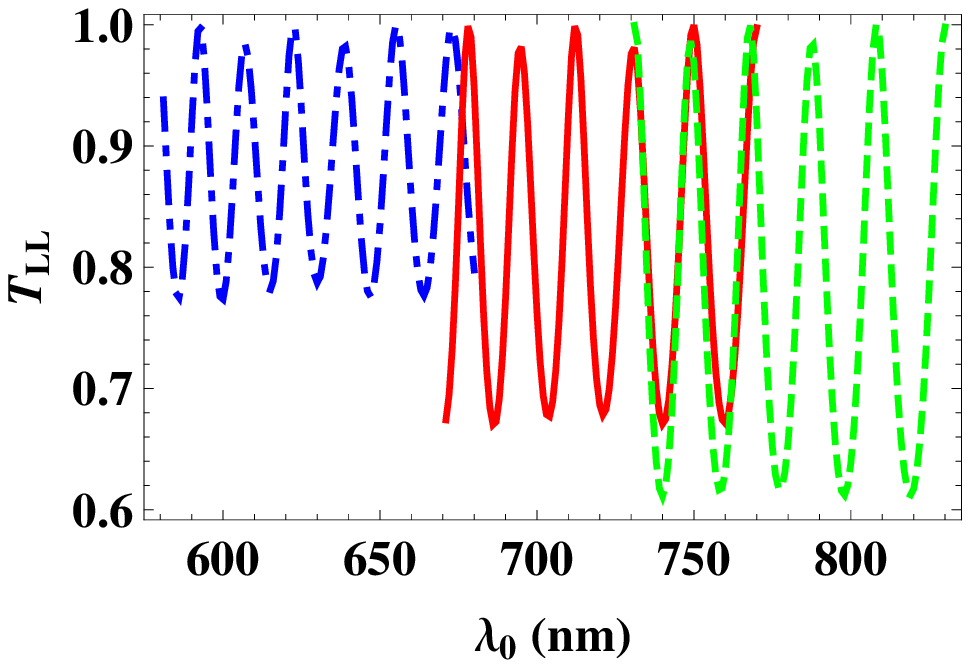} \hfill
\includegraphics[width=2.6in]{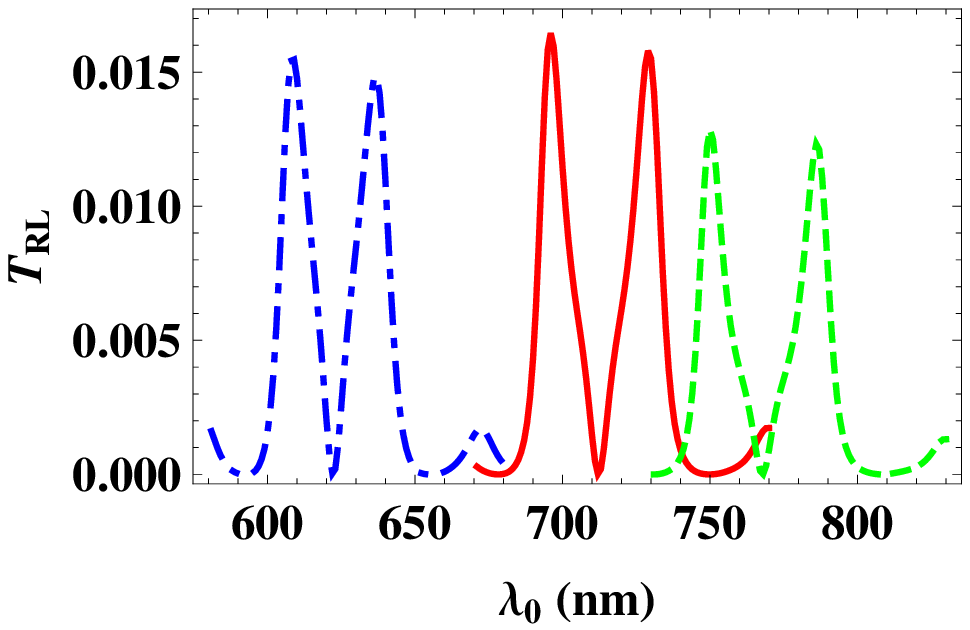}\\
\includegraphics[width=2.6in]{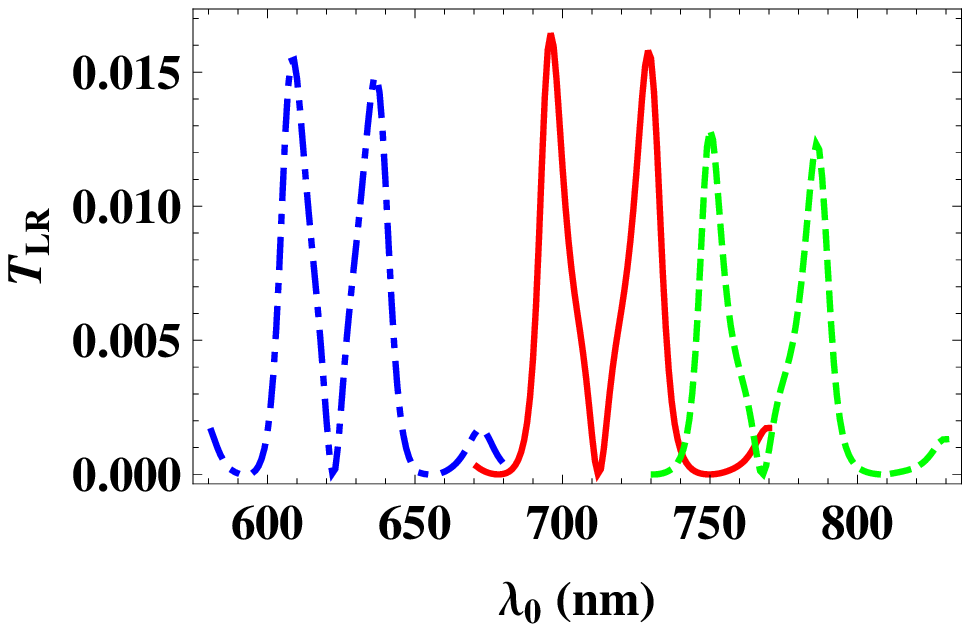} \hfill
\includegraphics[width=2.6in]{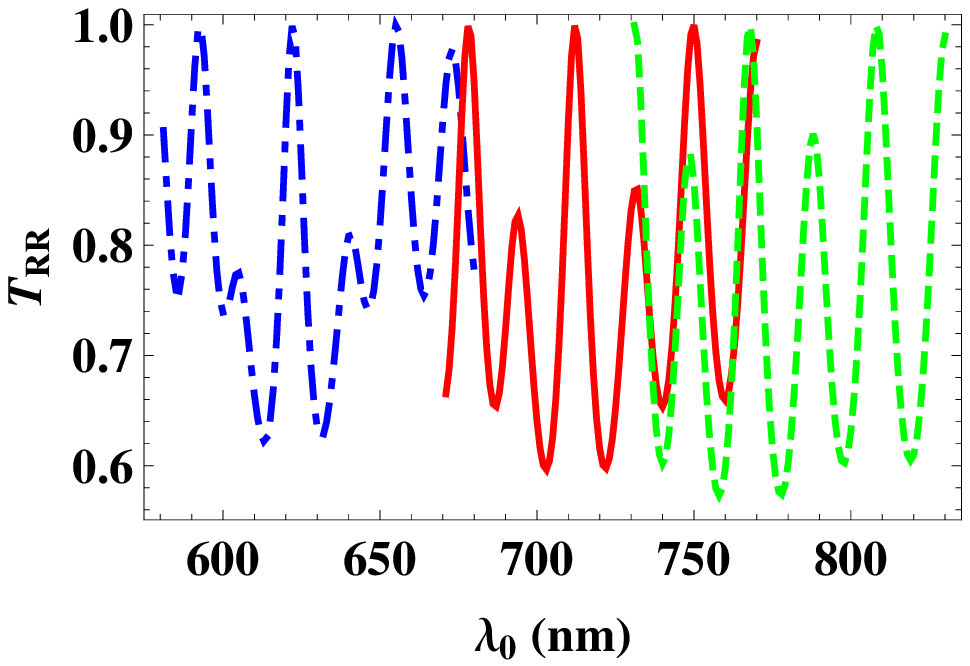}
 \caption{\l{Fig2}
 As Fig.~\ref{Fig1} except that the upper half $z \in \les  L/2, L \ris$  of the CSTF is twisted
 about the $z$ axis by $\pi/2$ radians with respect to the lower half $z \in \les 0, L/2 \ri$.
}
\end{figure}

\newpage
\begin{figure}[!h]
\centering
\includegraphics[width=3.3in]{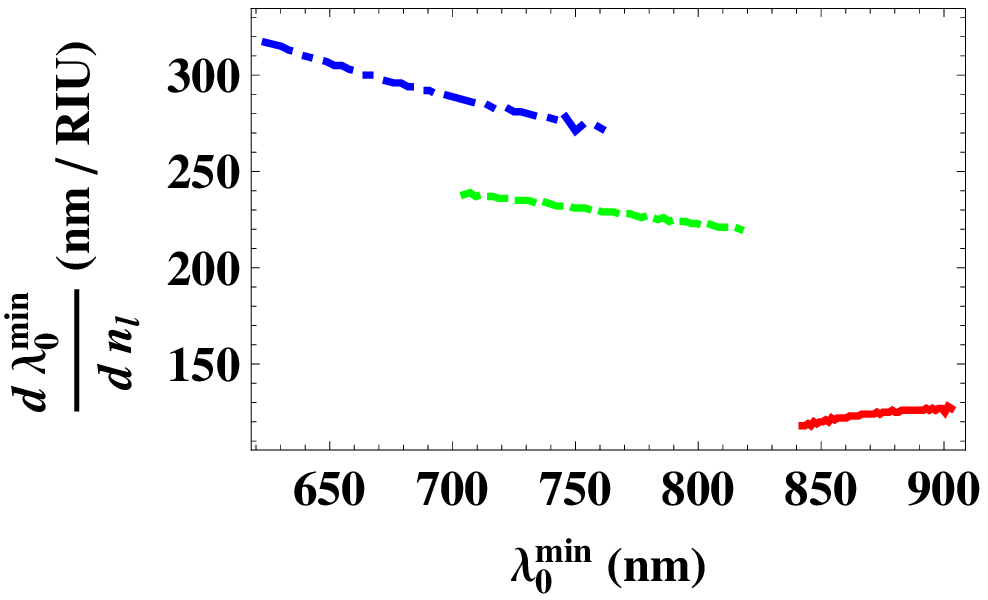}
 \caption{\l{Fig2a}
Spectral-shift sensitivity $d \lambda^{min}_0 / d n_\ell$
  plotted against $\lambda^{min}_0$ for $n_\ell \in \le 1, 1.5 \ri$.
 The vapor flux angle $\chi_v =
15^\circ$ (blue broken-dashed curve), $30^\circ$ (green dashed
curve), and $60^\circ$ (red solid curve). }
\end{figure}

\end{document}